\pdfoutput=1

\documentclass[a4paper,conference]{IEEEtran}

\usepackage[T1]{fontenc}
\usepackage[utf8]{inputenc}
\usepackage{csquotes}
\usepackage[noadjust]{cite}

\usepackage{amsmath}
\usepackage{amssymb}
\DeclareMathOperator{\diag}{diag}
\DeclareMathOperator{\blkdiag}{blkdiag}

\usepackage{mathtools}
\DeclarePairedDelimiter{\abs}{\lvert}{\rvert}
\DeclarePairedDelimiter{\norm}{\lVert}{\rVert}

\newbool{pdffigures}
\setbool{pdffigures}{true}

\DeclarePairedDelimiter{\parens}{\lparen}{\rparen}
\DeclarePairedDelimiter{\floor}{\lfloor}{\rfloor}
\DeclareMathOperator{\Real}{Re}
\DeclareMathOperator{\E}{E}
\DeclareMathOperator{\Imag}{Im}
\DeclareMathOperator{\tr}{tr}

\DeclareMathOperator{\SPEB}{SPEB}
\DeclareMathOperator{\PEB}{PEB}

\usepackage{url}
\usepackage{microtype}

\input{glyphtounicode-cmr.tex}
\input{glyphtounicode-ntx.tex}

\usepackage{bm}
\newcommand{\matt}[1]{\bm{#1}}
\newcommand{\vect}[1]{\bm{#1}}

\usepackage{hyperref}
\hypersetup{pdfauthor={Tobias Laas;Ronald Boehnke;Wen Xu},pdftitle={Optimal Beamforming for Bistatic MIMO Sensing},pdfkeywords={Bistatic sensing;OFDM sensing;MIMO sensing;optimal beamforming}}
\usepackage{doi}

\usepackage{tikz}

\usepackage{textcomp}
\usepackage{siunitx}

\usepackage{pgfplots}
\usetikzlibrary{backgrounds}
\usetikzlibrary{calc}
\tikzstyle{thicker}=[line width=1pt]

\newcommand{\inputexttikzfig}[1]{\ifbool{pdffigures}{\includegraphics{#1}}{\tikzsetnextfilename{#1}\input{#1}}}
\usepackage{afterpage}
\usepackage[outline]{contour}
\contourlength{0.8pt}

\newtheorem{statement}{Statement}
\DeclareSIUnit\decibelmW{dBm}
\begin{document}
	\allowdisplaybreaks[4]
\bstctlcite{BSTcontrol}
\title{Optimal Beamforming for Bistatic MIMO Sensing}
\author{\IEEEauthorblockN{Tobias Laas, Ronald Boehnke and Wen Xu}
	\IEEEauthorblockA{Munich Research Center, Huawei Technologies Duesseldorf GmbH, Munich, Germany\\
	Email: tobias.laas@tum.de; ronald.boehnke@huawei.com; wen.xu@ieee.org}}
\maketitle
\begin{tikzpicture}[remember picture,overlay]
\node[yshift=2cm] at (current page.south){\parbox{\textwidth}{\footnotesize \textcopyright\ 2024 IEEE. Personal use of this material is permitted. Permission from IEEE must be obtained for all other uses, in any current or future media, including reprinting/republishing this material for advertising or promotional purposes, creating new collective works, for resale or redistribution to servers or lists, or reuse of any copyrighted component of this work in other works.}};
\node[yshift=-1cm] at (current page.north){\parbox{\textwidth}{\footnotesize This is the accepted version of the following article: 
T.~Laas, R.~Boehnke, and W.~Xu, ``Optimal beamforming for bistatic {MIMO} sensing,'' in \emph{Proc. IEEE Global Commun. Conf. Workshops (GC Workshops)}, Cape Town, South Africa, Dec. 2024, \doi{10.1109/GCWkshp64532.2024.11101571}.}};
\end{tikzpicture}%
\begin{abstract}
    This paper considers the beamforming optimization for sensing a point-like scatterer using a bistatic multiple-input multiple-output (MIMO) orthogonal frequency-division multiplexing (OFDM) radar, which could be part of a joint communication and sensing system. The goal is to minimize the Cramér-Rao bound on the target position's estimation error, where the radar already knows an approximate position that is taken into account in the optimization. The optimization considers multiple subcarriers, and permits beamforming with more than one beam per subcarrier. We discuss the properties of optimal beamforming solutions, including the case of a known channel gain. Numerical results show that beamforming with at most one beam per subcarrier is optimal for certain parameters, but for other parameters, optimal solutions need two beams on some subcarriers. In addition, the degree of freedom which end of the bistatic radar should transmit and receive in a bidirectional radar is considered.
\end{abstract}
\section{Introduction}
\setlength{\lineskip}{1pt}
\setlength{\lineskiplimit}{1pt}
In this paper, we consider a bistatic multiple-input multiple-output (MIMO) orthogonal frequency-division multiplexing (OFDM) radar that is sensing a point-like target. The book~\cite{BistaticRadarBook} provides a nice overview on bistatic radars. One advantage of bistatic radars compared to monostatic radars is that the transmit/receive points (TRPs) do not need to support full-duplex transmission. There is a large body of literature, which only considers bistatic systems or networks with single transmit and receive antennas at each node, see e.g.,~\cite{Review1,Review2,Review3,Review4}. Bistatic radars are particularly interesting in the context of joint communications and sensing (JC\&S) in cellular systems with base station cooperation, because many cellular systems do not support full-duplex and because base stations can exchange information over the front-, mid- or backhaul links instead of needing valuable air interface resources. 

The beamforming in a MIMO radar, which was first proposed in~\cite{MIMORadar}, can improve the signal-to-noise ratio (SNR) and allows for spatial filtering at the same time. There are well-known methods to optimize it based on the transmit covariance matrix: e.g., the approximation of a given target beampattern~\cite{Stoica2007}. But such an optimization leads to different results compared to optimizing a bound on the estimation performance. We aim to optimize the beamforming and compare the estimation performance measured by the Cramér-Rao bound (CRB) on the target position's estimation error, where the transmitter already knows an approximate position. The optimization could be applied in a target tracking loop. In~\cite{AdhamJCnS2021}, the CRB for a multistatic radar based on the  azimuth angles of arrival (AoA) and departure (AoD) and the delay between transmitter and receiver is considered, but without an optimized beamforming. A similar CRB is obtained for single-anchor positioning in~\cite{KakkavasTWC2019}, which additionally includes the transform from AoA, AoD and delay to position. 
The present work was inspired by~\cite{UniBoICCWS2023}, where a single beam is optimized for position estimation based on the azimuth AoA and AoD in a single-frequency system / on a single subcarrier only, where delay estimation cannot be used and which requires the radar cross section (RCS) to be known.

\emph{Our contribution} is that we extend this by considering multiple subcarriers in an OFDM system, enabling the estimation of the delay between transmitter and receiver for the position estimation, and by permitting more than one beam per subcarrier, which has not been considered to the authors' best knowledge. With this approach, knowledge of the RCS is not necessary. We optimize the beamforming and derive the properties of optimal solutions, including the degree of freedom in selecting which end of the bistatic radar should transmit and receive in a bidirectional radar. Further, we consider the case of a known channel gain as a bound on how much the performance could be improved by channel gain tracking. In the following section, a system model based on the azimuth AoA and AoD and the delay between transmitter and receiver is presented, which is valid in the far-field. 

\textit{Notation}: lowercase bold letters denote vectors, uppercase
bold letters matrices. $\matt A^T$, $\matt A^\ast$, $\matt A^H$, $\norm{\vect a}$ and $\tr(\matt A)$ correspond to the transposed, the complex conjugate, the Hermitian, the Euclidean norm and the trace respectively. $\diag(d_1,\dotsc,d_N)$ denotes the diagonal matrix with $d_1$ to $d_N$ on its main diagonal, and $\blkdiag(\matt B_1,\dotsc,\matt B_N)$ the block-diagonal matrix with $\matt B_1$ to $\matt B_N$ as its main-diagonal blocks. $\vect 0$ and $\matt I$ are the zero vector/matrix and identity matrix. $\mathcal{N}_\mathbb{C}(\mu, \sigma^2)$ denotes a circularly-symmetric complex Gaussian distribution with mean $\mu$ and variance $\sigma^2$, and $\E[ X ]$ the expectation of the random variable $X$.

\section{System Model}
\begin{figure}[t]
	\centering
	\input{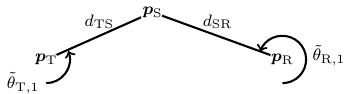}%
	\caption{Geometry for bistatic sensing.}
	\label{fig:bistaticgeometry}
\end{figure}%
The following model is based on the framework presented in~\cite{KakkavasTWC2019}.
Consider a bistatic MIMO OFDM radar supporting $P$ subcarriers with a transmitter with $N_\mathrm{T}$ antennas located at the 2D position $\vect p_\mathrm{T}$ and a receiver with $N_\mathrm{R}$ antennas located at $\vect p_\mathrm{R}$ sensing a target at the position $\xmathstrut[0]{0.65}\vect p_\mathrm{S}^{T} = \begin{bmatrix}x_\mathrm{S}, y_\mathrm{S}\end{bmatrix}$, see Fig.~\ref{fig:bistaticgeometry}. $d_\mathrm{TS}$ and $d_\mathrm{SR}$ are the distances between the scatterer and the transmitter, and the scatterer and the receiver. 

The received signal on the $p$-th subcarrier $\vect y[p]  \in \mathbb{C}^{N_\mathrm{R}}$ in an additive white Gaussian noise (AWGN) channel is given by
\begin{gather}
	\label{equ:sysmodel}
	\vect y[p]=\vect m[p]+\vect \eta[p],\quad \vect \eta[p] \sim \mathcal{N}_\mathbb{C}(\vect 0,\sigma_\eta^2 \matt I),\\
	\label{equ:mp}
	\vect m[p]=\sum\limits_{l=0}^1 h_l \vect a_{\mathrm{R},l}[p](\tilde{\theta}_{\mathrm{R},l}) \vect a_{\mathrm{T},l}^T[p](\tilde{\theta}_{\mathrm{T},l}) e^{-j \omega_p \tau_l} \vect s[p],
\end{gather}
where $\sigma_\eta^2$ is the noise covariance, $h_l$ is the $l$-th path's complex channel coefficient, $\vect a_{\mathrm{R},l}, \vect a_{\mathrm{T},l}$ are the array response vectors, $\omega_p$ is the angular frequency of the $p$-th subcarrier in the baseband, $\tau_l$ is the delay on the $l$-th path, and $\tilde{\theta}_{\mathrm{R},l}$ and $\tilde{\theta}_{\mathrm{T},l}$ are the AoA and AoD\@.  Note that $\vect s[p] \in \mathbb{C}^{N_\mathrm{T}}$ and $\vect y[p]$ correspond to the signals before spreading and after despreading respectively.

In the remainder of the paper, we only consider the path $l=1$ in the estimation of the scatterer's position $\vect p_\mathrm{S}$, because the line-of-sight path ($l=0$) does not give any information for this estimation, since the position and orientation of the transmitter and the receiver are assumed to be known, and because we assume that clutter has been removed, i.e. we focus on the target position estimation. Further, we assume that $\vect a_{\mathrm{R},1}$ and $\dot{\vect a}_{\mathrm{R},1}$, and $\vect a_{\mathrm{T},1}$ and $\dot{\vect a}_{\mathrm{T},1}$ are orthogonal
by choice of a suitable phase center~\cite[Sect.~A.1.1]{ClassicalDoACh1}, or in other words a suitable local coordinate system, where
\begin{equation}
	\dot{\vect a}_{\mathrm{R},l} = \partial \vect a_{\mathrm{R},l} / \partial \tilde{\theta}_{\mathrm{R},l} ,\qquad\dot{\vect a}_{\mathrm{T},l} = \partial \vect a_{\mathrm{T},l}/\partial\tilde{\theta}_{\mathrm{T},l} .
\end{equation}
Let $h_{1,\mathrm{R}} = \Real(h_1)$, $h_{1,\mathrm{I}} = \Imag(h_1)$. $h_1$ is an unknown parameter, whose phase is not only influenced by the channel's phase, but also by the transmitter and receiver not being phase synchronized. The parameter vector for the estimation of the scatterer's position is given by 
\begin{equation}
	\vect \phi_\mathrm{S}^T = \begin{bmatrix} h_{1,\mathrm{R}},h_{1,\mathrm{I}},x_\mathrm{S},y_\mathrm{S}\end{bmatrix}.
\end{equation}
Instead of directly estimating the parameters based on $\vect \phi_\mathrm{S}$, we consider the parameter vector based on angles and delay
\begin{equation}
	\vect \phi^T =  \begin{bmatrix}\phi_1,\phi_2,\phi_3,\phi_4,\phi_5\end{bmatrix}=\begin{bmatrix} h_{1,\mathrm{R}},h_{1,\mathrm{I}},\tau_1,\tilde{\theta}_{\mathrm{T},1},\tilde{\theta}_{\mathrm{R},1}\end{bmatrix},
\end{equation}
since our system model, see \eqref{equ:sysmodel} and \eqref{equ:mp}, is parameterized by them.
The entries of the corresponding Fisher information matrix (FIM) $\matt J \in \mathbb{R}^{5\times 5}$ are given by
\begin{equation}
	\label{equ:defFI}
	J_{a,b} = \frac{2}{\sigma_\eta^2} \sum\limits_p\Real\left(\E\left[\frac{\partial \vect m[p]}{\partial \phi_a}^H\frac{\partial \vect m[p]}{\partial \phi_b}\right]\right),
\end{equation}
see~\cite[Sect.~15.7]{Kay1993}. The derivatives can be found in the Appendix. $\matt J$ can be partitioned in the following way
\begin{equation}
	\matt J = \begin{bmatrix}\matt J_{11} & \matt J_{12}\\
\matt J_{12}^T &  \matt J_{22} \end{bmatrix}, \quad \matt J_{11} \in \mathbb{R}^{2\times2}, \matt J_{12} \in \mathbb{R}^{2\times 3}, \matt J_{22} \in \mathbb{R}^{3\times3}
\end{equation}
for the estimation of $\vect p_\mathrm{S}$.
The squared position error bound (SPEB) on $\vect p_\mathrm{S}$ is given by~\cite{MoePart1}
\enlargethispage{-0.123cm}
\begin{multline}
	\label{equ:SPEB}
	\SPEB = \tr\parens[\Big]{\left(\matt K \matt J_\mathrm{e} \matt K^T\right)^{-1}},\quad \matt J_\mathrm{e} = \matt J_{22} - \matt J_{12}^T \matt J_{11}^{-1} \matt J_{12},\\
	\xmathstrut[0]{0.95}\matt K = \begin{bmatrix} \frac{\partial \tau_1}{\partial \vect p_\mathrm{S}} & \frac{\partial \tilde{\theta}_{\mathrm{T},1}}{\partial \vect p_\mathrm{S}} & \frac{\partial \tilde{\theta}_{\mathrm{R},1}}{\partial \vect p_\mathrm{S}}\end{bmatrix},
\end{multline}
where $\matt J_\mathrm{e}$ is the equivalent FIM (EFIM) on $[\tau_1,\tilde{\theta}_{\mathrm{T},1},\tilde{\theta}_{\mathrm{R},1}]$. It takes into account the reduced information due to the unknown channel coefficient $h_1$. For a sufficiently high SNR and sufficient prior information, the bound can be achieved by the maximum likelihood estimator.
Computing the derivatives in $\matt K$ gives
\begin{equation}
	\label{equ:Kval}
	\matt K = \begin{bmatrix} \frac{\vect e_r(\theta_{\mathrm{T},1})+\vect e_r(\theta_{\mathrm{R},1})}{c} &  \frac{\vect e_\varphi(\theta_{\mathrm{T},1})}{d_\mathrm{TS}} & \frac{\vect e_\varphi(\theta_{\mathrm{R},1})}{d_\mathrm{SR}} \end{bmatrix},
\end{equation}
\enlargethispage{-0.058cm}
where $c$ is the speed of light, and $\vect e_r(\varphi)$ and $\vect e_\varphi(\varphi)$ are the usual unit vectors in the polar coordinate system.
Evaluating \eqref{equ:defFI} gives the following structure for $\matt J_{22}$:
\begin{align}
	\matt J_{22} &= \begin{bmatrix} j_{33} & j_{34} & 0 \\
		j_{34} & j_{44} & 0 \\
		0 & 0 & j_{55}\end{bmatrix},\quad \matt R_{s}[p]=\E[\vect s[p]\vect s[p]^H],\\
	\label{equ:j33}
	j_{33}&=\frac{2}{\sigma_\eta^2}  N_\mathrm{R}  \abs{h_1}^2 \sum\limits_p \omega_p^2 \vect a_{\mathrm{T},1}^T[p] \matt R_{s}[p] \vect a_{\mathrm{T},1}^\ast[p],\\
	\label{equ:j44}
	j_{44}&=\frac{2}{\sigma_\eta^2} N_\mathrm{R} \abs{h_1}^2 \sum\limits_p \dot{\vect a}_{\mathrm{T},1}^T[p] \matt R_{s}[p] \dot{\vect a}_{\mathrm{T},1}^\ast[p],\\
	\label{equ:j55}
	j_{55}&=\frac{2}{\sigma_\eta^2} \abs{h_1}^2 \sum\limits_p \norm{\dot{\vect a}_{\mathrm{R},1}[p]}^2\vect a_{\mathrm{T},1}^T[p] \matt R_{s}[p] \vect a_{\mathrm{T},1}^\ast[p],\\
	j_{34} &= -\frac{2}{\sigma_\eta^2} N_\mathrm{R} \abs{h_1}^2 \sum\limits_p\omega_p  \Imag(\dot{\vect a}_{\mathrm{T},1}^T[p] \matt R_s[p] \vect a_{\mathrm{T},1}^\ast[p]).
\end{align}
For the EFIM, we also need the subtrahend
\begin{gather}
	\setlength\arraycolsep{4.55pt}
\nonumber	\matt J_{12}^T \matt J_{11}^{-1} \matt J_{12} = \frac{1}{j_{11}} \begin{bmatrix}j_{13}^2+j_{23}^2 & j_{13}j_{14}+j_{23}j_{24} & 0\\
		            j_{13}j_{14}+j_{23}j_{24} & j_{14}^2 + j_{24}^2 & 0 \\
		            0 & 0 & 0\end{bmatrix},\\
   \label{equ:j33schur}
	\frac{j_{13}^2+j_{23}^2}{j_{11}}= \frac{2}{\sigma_\eta^2} N_\mathrm{R} \abs{h_1}^2 \frac{\bigl(\sum_p\omega_p  \vect a_{\mathrm{T},1}^T \matt R_s[p] \vect a_{\mathrm{T},1}^\ast\bigr)^2}{\sum_p \vect a_{\mathrm{T},1}^T \matt R_s[p] \vect a_{\mathrm{T},1}^\ast},\\
	\label{equ:j44schur}
	\frac{j_{14}^2+j_{24}^2}{j_{11}}= \frac{2}{\sigma_\eta^2} N_\mathrm{R} \abs{h_1}^2 \frac{\abs[\big]{\sum_p \dot{\vect a}_{\mathrm{T},1}^T \matt R_s[p] \vect a_{\mathrm{T},1}^\ast}^2}{\sum_p \vect a_{\mathrm{T},1}^T\matt R_s[p] \vect a_{\mathrm{T},1}^\ast},\\
	\label{equ:j1313232411}
\begin{split}&\frac{j_{13} j_{14}+j_{23} j_{24}}{j_{11}} = \frac{-2}{\sigma_\eta^2} N_\mathrm{R} \abs{h_1}^2 \\&\cdot\frac{\bigl(\sum_p\omega_p  \vect a_{\mathrm{T},1}^T \matt R_s[p] \vect a_{\mathrm{T},1}^\ast\bigr)\bigl(\sum_p \Imag(\dot{\vect a}_{\mathrm{T},1}^T \matt R_s[p] \vect a_{\mathrm{T},1}^\ast)\bigr)}{\sum_p \vect a_{\mathrm{T},1}^T \matt R_s[p] \vect a_{\mathrm{T},1}^\ast}.\end{split}
\end{gather}
The entries of $\matt J_{11}$ and $\matt J_{12}$ can be found in the Appendix. According to \eqref{equ:j33} -- \eqref{equ:j1313232411}, the EFIM only depends on the transmission into the directions $\vect a_{\mathrm{T},1}^T[p]$ and $\dot{\vect a}_{\mathrm{T},1}^T[p]$. For a given transmit power $P_\mathrm{T}$, it is optimal to choose the precoder $\matt F_p$ such that
\begin{equation}
	\label{equ:precoding}
\matt R_s[p] = \matt F_p \matt B_p \matt F_p^H, \quad \matt F_p =\begin{bmatrix} \frac{\vect a_{\mathrm{T},1}^\ast[p]}{\norm{\vect a_{\mathrm{T},1}[p]}} & \frac{\dot{\vect a}_{\mathrm{T},1}^\ast[p]}{\norm{\dot{\vect a}_{\mathrm{T},1}[p]}}\end{bmatrix},
\end{equation}
where $\matt B_p$ is a Hermitian positive semi-definite covariance matrix. It is suboptimal to transmit into other directions, because they do not contribute to the EFIM and thus transmit power is wasted. The transmission into the direction $\vect a_{\mathrm{T},1}^T[p]$ is required for AoA and delay estimation, while the transmission into the direction $\dot{\vect a}_{\mathrm{T},1}^T[p]$ is required for AoD estimation. 

\section{Beamforming Optimization}
Consider the minimization of the $\SPEB$ via beamforming for a symmetric multicarrier system ($\omega_p = -\omega_{-p}\ \forall p$) w.l.o.g.\@
\begin{equation} 
	\label{equ:optim}
	\min_{\matt B \succeq \matt 0, \tr(\matt B)\le P_\mathrm{T}} \tr\left(\left(\matt K \matt J_\mathrm{e}(\matt B) \matt K^T\right)^{-1}\right),
\end{equation}
where $\matt B = \blkdiag(\matt B_{-P_\mathrm{m}}, \dotsc, \matt B_{P_\mathrm{m}})$, and $P_{\mathrm{m}}=\floor{P/2}$ is the index of the subcarrier with the highest frequency. Note that the system needs to know an approximate position of the target in the beamforming optimization, which is needed to compute the $\SPEB$.
\begin{statement}
The optimization problem is convex in $\matt B$.
\end{statement}
\begin{IEEEproof}
	Let us re-write \eqref{equ:SPEB} based on $\matt J$ instead of $\matt J_\mathrm{e}$ using the properties of the Schur-complement:
\begin{equation}
	\begin{split}
	\SPEB &= \tr(\matt U (\matt K_1 \matt J \matt K_1^T)^{-1} \matt U^T),\\
	\matt U&=\begin{bmatrix}\vect u_1^H\\\vect u_2^H\end{bmatrix}=\begin{bmatrix}\matt 0 & \matt I\end{bmatrix}\in\mathbb{R}^{2\times 4}, \quad \matt K_1 = \begin{bmatrix} \matt I & \matt 0\\\matt 0 & \matt K\end{bmatrix}.
	\end{split}
\end{equation}
Based on the precoding \eqref{equ:precoding}, $\matt J$ can be written as
\begin{equation}
	\begin{split}
	&\matt J = \sum\limits_p \frac{2}{\sigma_\eta^2} \Real(\abs{h_1}^2\matt X_{\mathrm{R},p} \matt B_p \matt X_{\mathrm{R},p}^H + \matt X_{\mathrm{T},p} \matt B_p \matt X_{\mathrm{T},p}^H),\\
	&\matt X_{\mathrm{R},p} = \sqrt{N_\mathrm{T}}\begin{bmatrix}\vect 0 & \vect 0\\ \norm{\dot{\vect a}_{\mathrm{R},1}[p]} & 0\end{bmatrix} \in \mathbb{R}^{5\times 2},\\
	&\matt X_{\mathrm{T},p} = \sqrt{N_\mathrm{R}}\begin{bmatrix*}[r]\sqrt{N_\mathrm{T}} & 0\\
		j \sqrt{N_\mathrm{T}} & 0\\
		-j h_1 \omega_p \sqrt{N_\mathrm{T}} & 0\\
		0 & h_1\norm{\dot{\vect a}_{\mathrm{T},1}[p]}\\
		0 & 0\end{bmatrix*},
\end{split}
\end{equation}
after solving several sets of equations.
After vectorization of the sum,
\begin{equation}
	\matt J = 2/\sigma_\eta^2  \cdot \Real(\abs{h_1}^2\matt X_\mathrm{R}\matt B \matt X_\mathrm{R}^H + \matt X_\mathrm{T}\matt B \matt X_\mathrm{T}^H).
\end{equation}
Since $\matt K_1 \matt J \matt K_1^T$ is positive definite for any feasible $\matt B$, and $\vect u^H \matt Y^{-1} \vect u$ is convex in $\matt Y$ for any $\vect u \in \mathbb{R}^4$ and $\matt Y \succ \matt 0 \in \mathbb{R}^{4 \times 4}$, see~\cite[Sect.~1]{ConvexityInverse}, $\SPEB(\matt B)$ is convex, because $\matt K_1 \matt J \matt K_1^T$ is linear in $\matt B$ and the $\tr$ can be expanded: $\tr(\matt U \matt Y^{-1} \matt U) = \vect u_1^H \matt Y^{-1} \vect u_1 +\vect u_2^H \matt Y^{-1} \vect u_2$. It can be solved using a projected gradient method similarly to~\cite{HungerDissBook} for example.
\end{IEEEproof}

\section{Optimal Beamforming}
Consider the optimal diagonal entries of $\matt B_p$, $b_{p,11}$ and $b_{p,22}$, for a transmitter with $N_\mathrm{T}\ge 2$, because for $N_\mathrm{T}=1$ there is no beamforming.
In a narrowband system, $\vect a_{\mathrm{T},l}[p]$, $\vect a_{\mathrm{R},l}[p]$, $\dot{\vect a}_{\mathrm{T},l}[p]$ and $\dot{\vect a}_{\mathrm{R},l}[p]$ are the same for all $p$. 

Let us first consider the delay and the AoA and AoD estimation separately:
\begin{itemize}
	\item The delay estimation depends on the power allocated to $b_{p,11}$ on the subcarriers regardless of the narrowband assumption, because the summands in \eqref{equ:j33} are weighted by $\omega_p^2$. It is well-known from time of arrival estimation that it is CRB-optimal to allocate all power to the outermost subcarriers $p=\pm P_\mathrm{m}$~\cite{Xu2018}. Delay estimation needs at least two subcarriers.
	\item With the narrowband assumption, the equivalent Fisher information corresponding to AoA and AoD estimation depends on the sum power transmitted towards $\vect a_{\mathrm{T},1}^T$ and $\dot{\vect a}_{\mathrm{T},1}^T$ over all subcarriers, and is independent of how it is allocated to the subcarriers, because $b_{p,11}$ and $b_{p,22}$ only appear in the sums $\sum_p b_{p,11}$ and $\sum_p b_{p,22}$ in \eqref{equ:j44}, \eqref{equ:j55} and \eqref{equ:j44schur}. 
Without this assumption however, $j_{44}$ and $j_{55}$ do depend on $\norm{\dot{\vect a}_{\mathrm{T},1}[p]}$ and $\norm{\dot{\vect a}_{\mathrm{R},1}[p]}$ respectively, which increase with the subcarrier index. This means that it is beneficial to allocate more power to the higher subcarriers.
\end{itemize}
Second, let us consider the impact of $\Real(b_{p,21})$: the EFIM only depends on it via $(j_{14}^2+j_{24}^2)/j_{11}$ in \eqref{equ:j44schur}. The $\SPEB$ and the Fisher information are minimized and maximized respectively by $\Real(b_{p,21})=0$, because the constants in front of the fraction, as well as the numerator and the denominator are positive. This means that $\Real(b_{p,21})= 0$ holds for all $p$ in an optimal solution.

Third, let us put these considerations together: the condition $\matt J_{12}^T \matt J_{11}^{-1} \matt J_{12} = \matt 0$, such that there is no loss in information due to the unknown $h_1$, is equivalent to $\sum_p \omega_p b_{p,11}=0$ and $\sum_p \norm{\dot{\vect a}_{\mathrm{T},1}[p]}b_{p,21} = 0$. This condition is not fulfilled in an optimum in general, because there is a trade-off between minimizing $\abs{\sum_p \omega_p b_{p,11}}$ for the condition to hold and maximizing $\sum_p \norm{\dot{\vect a}_{\mathrm{R},1}[p]}^2 b_{p,11}$ for improving AoA estimation, see \eqref{equ:j55}. With the narrowband assumption however, there is no trade-off due to the symmetry of the problem and the symmetric solution $b_{P_\mathrm{m},11}=b_{-P_\mathrm{m},11}>0$, $b_{p,11}=0\ \forall p \ne \pm P_\mathrm{m}$, $\Imag(b_{p,21})=-\Imag(b_{-p,21})$ is optimal, and $b_{p,22}$ can be allocated arbitrarily for a given $\sum_p b_{p,22}$ as long as all $\matt B_p$ are positive semidefinite.

There is an interesting solution that is optimal in some cases, see Sect.~\ref{sec:numres}: $\matt B_p = \matt 0$ for $p \ne \pm P_{\mathrm{m}}$ and
\begin{equation}
	\matt B_{\pm P_{\mathrm{m}}} = \begin{bmatrix} \alpha & \pm j \sqrt{\alpha (1-\alpha)}\\
		\mp j \sqrt{\alpha (1-\alpha)} & 1-\alpha\end{bmatrix}\frac{P_\mathrm{T}}{2},
\end{equation}
for some $\alpha \in ]0,1[$. Note that $\matt B_{\pm P_{\mathrm{m}}}$ is rank-1 and corresponds to a transmit signal, where the beamformer on subcarrier $P_{\mathrm{m}}$ is tilted away from $\vect a_{\mathrm{T},1}^T$ into the direction $-\dot{\vect a}_{\mathrm{T},1}^T$ and into the direction $\dot{\vect a}_{\mathrm{T},1}^T$ on subcarrier $-P_{\mathrm{m}}$, i.e.,
\begin{multline}
	\vect s[\pm P_{\mathrm{m}}] = \left(\frac{\sqrt{\alpha}\vect a_{\mathrm{T},1}^\ast[\pm P_{\mathrm{m}}]}{\norm{\vect a_{\mathrm{T},1}[\pm P_{\mathrm{m}}]}} \mp j  \frac{\sqrt{1-\alpha}\dot{\vect a}_{\mathrm{T},1}^\ast[\pm P_{\mathrm{m}}]}{\norm{\dot{\vect a}_{\mathrm{T},1}[\pm P_{\mathrm{m}}]}} \right)\\\cdot x[\pm P_{\mathrm{m}}],\quad
	\E[x[\pm P_{\mathrm{m}}]] = 0, \quad \E[\abs{x[\pm P_{\mathrm{m}}]}^2] = \frac{P_\mathrm{T}}{2}.
\end{multline}
This is similar to a monopulse radar~\cite[Ch.~1]{MonopulseBook}, where two beams are formed at reception instead of transmission. But here, two transmit beams are required to estimate $h_1$. 
A small variation in the AoD causes a phase variation in $\vect y[p]$ similarly to a variation in delay. Correspondingly, $\matt J_\mathrm{e}$ becomes rank-deficient, since is impossible to estimate AoD and delay at the same time. To estimate $\vect p_\mathrm{S}$ anyway, $N_\mathrm{R} \ge 2$ is required, because $\matt K \matt J_\mathrm{e} \matt K^T$ can still be full-rank with AoA estimation. 

Note that a full-rank $\matt B_p$ can be implemented by simultaneously sending two pilot signals on the subcarrier, or by time-sharing between two pilot signals on the subcarrier. The pilot signals can either be deterministic or a random signals with matching (sample) covariance matrix. 
This matches the result from~\cite{CaireFundamentalTradeoff} that the optimal sample covariance matrix is deterministic. The random signals could be communication signals in a JC\&S system.

\section{\texorpdfstring{With Known Gain of $h_1$}{With Known Gain of h₁}}
In this section, we consider the case that the gain $\abs{h_1}$ is known, a bound on how much the performance could be improved by channel gain tracking. This can be included in the optimization by the additional constraint
\begin{equation}
	f = \abs{h_1}^2 - h_{1,\mathrm{R}}^2 - h_{1,\mathrm{I}}^2 = 0
\end{equation}
with its gradient w.r.t.~$\begin{bmatrix}h_{1,\mathrm{R}} & h_{1,\mathrm{I}} \end{bmatrix}^T$ given by
\begin{equation}
	\vect f^T = -2 \begin{bmatrix} h_{1,\mathrm{R}} & h_{1,\mathrm{I}} \end{bmatrix}.
\end{equation}
The $\SPEB$ with constraint can be computed by projecting the FIM onto the subspace orthogonal to $\vect f$, see~\cite{Stoica1998},
\begin{gather}
	\begin{split}
		\label{equ:SPEBconstr}
		&\SPEB = \tr(\matt U_1 (\matt K_2 \matt J \matt K_2^T)^{-1} \matt U_1^T),\\&\matt K_2 = \begin{bmatrix} \vect u^T & \vect 0^T\\\matt 0 & \matt K\end{bmatrix},\quad\matt U_1=\begin{bmatrix}\matt 0 & \matt I\end{bmatrix}\in\mathbb{R}^{2\times 3},
	\end{split}\\
	\vect u^T = \abs{h_1}^{-1} \begin{bmatrix} -h_{1,\mathrm{I}} & h_{1,\mathrm{R}} \end{bmatrix},\quad \norm{\vect u}=1, \quad \vect u^T \vect f = 0.
\end{gather}
Let us re-write \eqref{equ:SPEBconstr} in terms of the EFIM by use of the properties of the Schur-complement:
\begin{gather}
	\begin{split}
		&\SPEB = \tr((\matt K \matt J_\mathrm{eh} \matt K^T)^{-1}),\\
		&\matt J_\mathrm{eh} = \matt J_{22} - \matt J_{12}^T \vect u \left(\vect u^T \matt J_{11} \vect u\right)^{-1} \vect u^T \matt J_{12},
	\end{split}\\
	\begin{split}
		&\matt J_{12}^T \vect u \left(\vect u^T \matt J_{11} \vect u\right)^{-1} \vect u^T \matt J_{12}\\ &= \frac{1}{j_{11}} \begin{bmatrix}j_{13}^2+j_{23}^2 & j_{13}j_{14}+j_{23}j_{24} & 0\\
			j_{13}j_{14}+j_{23}j_{24} & j_\mathrm{eh} & 0 \\
			0 & 0 & 0\end{bmatrix},
	\end{split}\\
	\frac{j_\mathrm{eh}}{j_{11}}=\frac{2}{\sigma_\eta^2} N_\mathrm{R} \abs{h_1}^2 \frac{\bigl(\sum_p \Imag(\dot{\vect a}_{\mathrm{T},1}^T \matt R_s[p] \vect a_{\mathrm{T},1}^\ast)\bigr)^2}{\sum_p \vect a_{\mathrm{T},1}^T \matt R_s[p] \vect a_{\mathrm{T},1}^\ast}.
\end{gather}
Note that $j_\mathrm{eh}/j_{11}$ is the same as $(j_{14}^2+j_{24}^2)/j_{11}$ in \eqref{equ:j44schur} with the corresponding real part squared removed from the absolute value squared in the numerator. As the constants in front of the fraction, as well as the numerator and the denominator are positive, the receiver can have more information about the AoD if the channel gain is known, but only if there is at least one $\Real(b_{p,21})\ne 0$. There is no change for the other parameters. Therefore, there is no benefit in knowing the channel gain if an optimal beamforming is used, i.e. there is no benefit in channel gain tracking in our model. 
\enlargethispage{-0.057cm}
\section{Numerical Results}
\label{sec:numres}
\subsection{Fixed Transmitter and Receiver Role}
\label{subsec:fixed}
\begin{figure}[!t]
	\centering
	\input{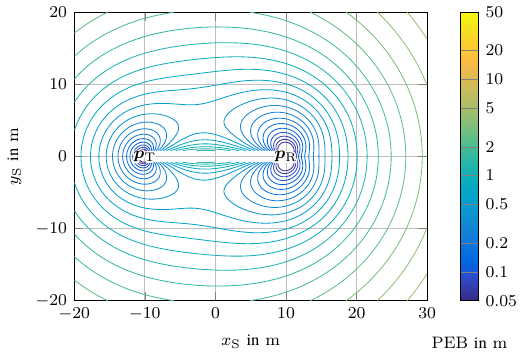}
	\caption{Logarithmic plot of the PEB for a varying $\vect p_\mathrm{S}$ with $P=2$ subcarriers and $N_\mathrm{T}=15$, $N_\mathrm{R}=3$.}
	\label{fig:PEB}
\end{figure}

Consider a symmetric multicarrier system at \SI{3.8}{\GHz} center frequency with $P=2$ subcarriers under the narrowband assumption, $N_\mathrm{T}=15$, $N_\mathrm{R}=3$ and uniform circular arrays (UCAs) with $\lambda/2$ antenna spacing. Let $\vect p_\mathrm{T}^T=[-10,0]\,\si{\meter}$, $\vect p_\mathrm{R}^T=[10,0]\,\si{\meter}$, $\sigma_\eta^2 = \SI{2.4e-14}{\watt}$, $P_\mathrm{T}=\SI{10}{\milli\W}$ and $\omega_p=2\pi p \cdot \SI{2.4}{\MHz}$, which corresponds to a noise spectral density of $\SI{-170}{\decibelmW/\hertz}$. The SPEB is independent of the phase of $h_1$, whose absolute value is modeled as
\begin{equation}
	\label{equ:absh1}
	\abs{h_1}=\SI{0.1}{\meter}\cdot \lambda/(4 \pi d_\mathrm{TS} d_\mathrm{SR}).
\end{equation}
The RCS in \eqref{equ:absh1} is constant: $\SI{0.01}{\meter^2}(4\pi)$. Note that the SPEB is independent of whether the path-loss is taken into account here, because the RCS is assumed to be unknown.
In Fig.~\ref{fig:PEB}, the position of the scatterer is varied, and the beamforming optimization is carried out for each grid point to obtain the position error bound ($\PEB = \sqrt{\SPEB}$). As expected, the $\PEB$ is smallest close to $\vect p_\mathrm{T}$ and $\vect p_\mathrm{R}$ and increases with increasing distance, or when the scatterer is close to the baseline of the radar, i.e.\ the line segment between $\vect p_\mathrm{T}$ and $\vect p_\mathrm{R}$~\cite[Ch.~3]{BistaticRadarBook}, because the delay and the angles give little information in this area. There are two regions for the contour lines: \begin{itemize} \item For a larger $\PEB$, there is an oval corresponding to a large $d_\mathrm{TS}$ and $d_\mathrm{SR}$, and there are two contour lines close to the baseline, one on each side of it. The oval resembles the well-known Cassini oval. 
	\item For a smaller $\PEB$, there are two contour lines, one around $\vect p_\mathrm{T}$ and one around $\vect p_\mathrm{R}$, similar to the Cassini ovals.
\end{itemize}

\begin{figure}[t]
	\centering
	\input{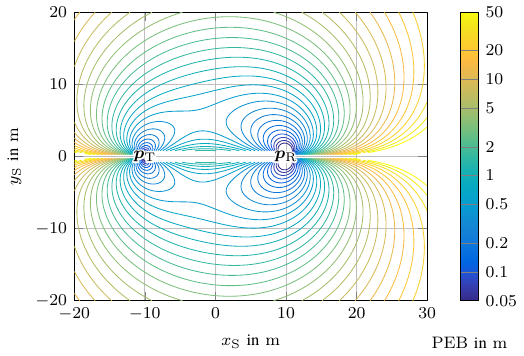}
	\caption{Logarithmic plot of the PEB for a varying $\vect p_\mathrm{S}$ with $P=1$ subcarrier and $N_\mathrm{T}=15$, $N_\mathrm{R}=3$.}
	\label{fig:bistaticbaseline}
\end{figure}

Let us compare this to a scenario inspired by~\cite{UniBoICCWS2023}, where we only have $P=1$ subcarrier, but the same transmit power, see Fig.~\ref{fig:bistaticbaseline}. Unlike~\cite{UniBoICCWS2023}, we assume that the RCS is unknown, which requires rank-2 beamforming on the subcarrier to ensure that $\matt K \matt J_\mathrm{e} \matt K^T$ is full-rank, because delay estimation is impossible with $P=1$, whereas only rank-1 beamforming is used in~\cite{UniBoICCWS2023}. The PEB obtained by optimization for $P=1$ is significantly larger than the PEB obtained with $P=2$ (Fig.~\ref{fig:PEB}), especially close to the half-lines that extend the baseline or further away from the baseline. Note that for $P=1$, it is impossible to estimate a $\vect p_\mathrm{S}$ that lies on the extended baseline, because the distance cannot be determined from the AoD and AoA due to the geometry, and in the vicinity, the geometry leads to a large PEB. The delay estimation is also highly beneficial further away from the baseline, because its equivalent Fisher information does not decrease as $d_\mathrm{TS}$ and $d_\mathrm{SR}$ increase, see \eqref{equ:Kval}, unlike the AoD's and AoA's. This significant performance difference between $P=1$ and $P=2$ shows that delay estimation is highly beneficial even at a small bandwidth ($\omega_{\pm 1} = \pm 2\pi \cdot\SI{2.4}{\mega\hertz}$).

\begin{figure}[!t]
	\centering
	\input{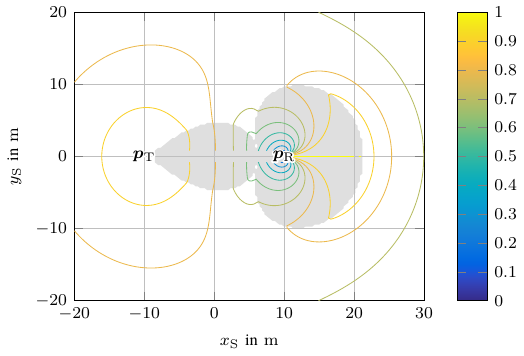}
	\caption{Share of $P_\mathrm{T}$ directed towards the scatterer, $\sum_p b_{p,11}/P_\mathrm{T}$, for $N_\mathrm{T}=15$, $N_\mathrm{R}=3$ and $P=2$. Rank-1 beamforming is optimal in the shaded area.}
	\label{fig:powerat}
\end{figure}%
\begin{figure}[!t]
	\centering
	\input{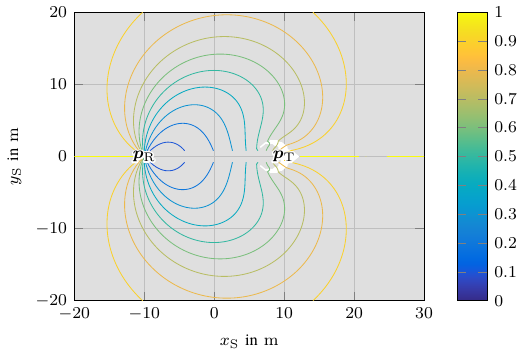}
	\caption{Share of $P_\mathrm{T}$ directed towards the scatterer, $\sum_p b_{p,11}/P_\mathrm{T}$, for $N_\mathrm{T}=3$, $N_\mathrm{R}=15$ and $P=2$. Rank-1 beamforming is optimal in the shaded area.}
	\label{fig:poweratswitched}
\end{figure}%
Let us return to the scenario with $P=2$: Fig.~\ref{fig:powerat} shows the share of transmit power allocated towards $\vect a_{\mathrm{T},1}^T$, i.e. $\sum_p b_{p,11}/P_\mathrm{T}$, which lies between $\SI{8.8}{\percent}$ and $\SI{100}{\percent}$ in the area considered. It is small close to $\vect p_\mathrm{R}$, but increases significantly on the side of $\vect p_\mathrm{R}$ that faces away from $\vect p_\mathrm{T}$ at the same time. As $d_\mathrm{TS}$ and $d_\mathrm{SR}$ increase, the information that the AoD and AoA give decreases, while that of the delay is independent of $d_\mathrm{TS}$ and $d_\mathrm{SR}$, see \eqref{equ:Kval}. Due to this, more power is allocated towards $\dot{\vect a}_{\mathrm{T},1}^T$ to compensate for the loss in AoD accuracy here, because $N_\mathrm{T}>N_\mathrm{R}$. 
For $N_\mathrm{R} \ge 2$, there are areas where rank-1 beamforming is optimal and where the full-rank solution is optimal, which is also shown in Fig.~\ref{fig:powerat}. The former is optimal, when the scatterer is close to the baseline or close to the receiver. In that part of the area close to $\vect p_\mathrm{T}$ where rank-1 beamforming is optimal, there is no benefit of a second beam, because delay estimation gives little information close to the baseline. In the corresponding area close to $\vect p_\mathrm{R}$, there is no benefit of a second beam, because the performance of AoA estimation is good, and AoD / delay estimation performs well. There is small notch between the two areas just discussed, where an additional beam that enables AoD and delay estimation is beneficial, because a large share of power is dedicated to AoD estimation, which reduces the information from AoA estimation, and delay estimation does not give much information, since the scatterer is close to the baseline.

Note that regardless of whether rank-1 or full rank beamforming are optimal, the optimal beams typically are weighted between $\vect a_{\mathrm{T},1}^T$ and $\dot{\vect a}_{\mathrm{T},1}^T$, i.e. $\Imag(b_{p,21})\ne 0$.

Consider also the power allocation and optimal strategy for the switched roles, i.e.\ $N_\mathrm{T}=3$, $N_\mathrm{R}=15$, see Fig.~\ref{fig:poweratswitched}. Due to the larger number of receive antennas, rank-1 beamforming is optimal almost everywhere. Similarly to Fig.~\ref{fig:powerat}, the share of power towards $\vect a_{\mathrm{T},1}^T$ is small close to $\vect p_\mathrm{R}$, but increases significantly on the side of $\vect p_\mathrm{R}$ that faces away from $\vect p_\mathrm{T}$ at the same time. But contrary to Fig.~\ref{fig:powerat}, the share of power allocated into $\dot{\vect a}_{\mathrm{T},1}^T$ decreases as $d_\mathrm{TS}$ and $d_\mathrm{SR}$ increase, because $N_\mathrm{R} > N_\mathrm{T}$ here.

\subsection{Switchable Transmitter and Receiver Role}
\begin{figure}[!t]
	\centering
	\ifbool{pdffigures}{
		\includegraphics{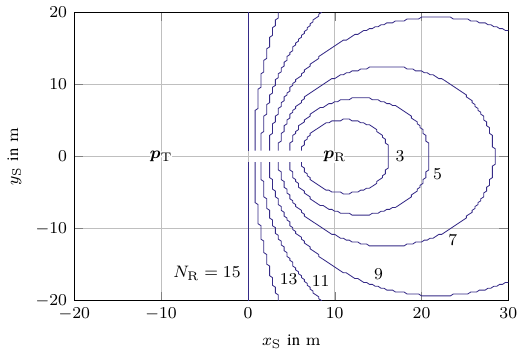}%
	}{
		\tikzsetnextfilename{bidi}
		\input{bidi_bidi}%
	}%
	\caption{Switchable roles for a varying $N_\mathrm{R}$ and a fixed $N_\mathrm{T}=15$, $P=2$: On the contour line(s), the performance of both directions is the same. For a scatterer outside of the contour line, it is better to switch roles.}
	\label{fig:bidi}
\end{figure}

There is an additional degree of freedom in a bidirectional bistatic radar system, where both ends can transmit and receive: one can select which TRP of the radar transmits and which receives. Fig.~\ref{fig:bidi} shows which TRP should transmit or receive based on the setup in Sect.~\ref{subsec:fixed}. In addition, $N_\mathrm{R}$ is varied from $3$ to $15$. Firstly, consider the case, where both TRPs are set up symmetrically, i.e.\ they have the same antenna arrays and symmetric orientation. In this case, the optimization result is that the TRP that is closer to the scatterer should always receive. Only when $d_\mathrm{TS}=d_\mathrm{SR}$, the performance in both directions is the same, and time-sharing between them is also optimal.

Secondly, consider $N_\mathrm{R}\ne 15$. Fig.~\ref{fig:bidi} shows that as in the symmetric setup, it is optimal that exactly one TRP transmits. The contour lines correspond to those scatterer positions with the same performance in both directions, where time-sharing is also optimal. As $N_\mathrm{R}$ increases, the area where the role of the TRPs shown in the figure is optimal increases.
Note that once it is known which TRP transmits, the optimal beamforming is the same as discussed in the previous subsection.

\section{Conclusion}
In this paper, the optimal beamforming for bistatic sensing for an OFDM system was discussed and it was shown numerically that a rank-1 solution is optimal for some parameters, and a full-rank solution is optimal for others, which is not considered by many papers in the literature. It was further shown that using more than one subcarrier is highly beneficial, because it enables delay estimation. Numerical results with the additional degree of freedom that both ends of the bistatic radar can transmit and receive in a bidirectional system show that it is optimal when exactly one TRP transmits, and one receives, while selecting which TRP should transmit and which receive varies with the number of antennas and the target's position.

\appendix[Intermediate Results Needed to Compute the FIM]
\label{sect:app}
The derivatives of $\vect m[p]$ w.r.t. the parameters included in the parameter vector $\vect \phi$ are given by
\begin{align}
	\frac{\partial \vect m[p]}{\partial h_{1,\mathrm{R}}}&=\vect a_{\mathrm{R},1}[p] \vect a_{\mathrm{T},l}^T[p] e^{-j \omega_p \tau_1} \vect s[p]=-j \frac{\partial \vect m[p]}{\partial h_{1,\mathrm{I}}},\\
	\frac{\partial \vect m[p]}{\partial \tau_1} &= -j \omega_p h_1 \vect a_{\mathrm{R},1}[p] \vect a_{\mathrm{T},1}^T[p] e^{-j \omega_p \tau_1} \vect s[p],\\
	\frac{\partial \vect m[p]}{\partial \tilde{\theta}_{\mathrm{T},1}} &= h_1 \vect a_{\mathrm{R},1}[p] \dot{\vect a}_{\mathrm{T},1}^T[p] e^{-j \omega_p \tau_1} \vect s[p],\\
	\frac{\partial \vect m[p]}{\partial \tilde{\theta}_{\mathrm{R},1}} &= h_1 \dot{\vect a}_{\mathrm{R},1}[p] \vect a_{\mathrm{T},1}^T[p] e^{-j \omega_p \tau_1} \vect s[p].
\end{align}
\newlength\lastpageshrink
\setlength\lastpageshrink{0cm}
\enlargethispage{-\lastpageshrink}
\afterpage{\enlargethispage{-\lastpageshrink}}
$\matt J_{11}$ and $\matt J_{12}$ are given by
\setlength\belowdisplayskip{0pt}
\begin{align}
	\matt J_{11} &= \diag(j_{11},j_{22}),\quad \matt J_{12} = \begin{bmatrix} j_{13} & j_{14} & 0\\j_{23} & j_{24} & 0\end{bmatrix},\\
	j_{11}&=j_{22} = 2/\sigma_\eta^2 \cdot N_\mathrm{R} \sum_p\vect a_{\mathrm{T},1}^T[p] \matt R_{s}[p] \vect a_{\mathrm{T},1}^\ast[p],\\
	j_{13} &= 2/\sigma_\eta^2 \cdot N_\mathrm{R} h_{1,\mathrm{I}} \sum_p\omega_p  \vect a_{\mathrm{T},1}^T[p] \matt R_s[p] \vect a_{\mathrm{T},1}^\ast[p],\\
	j_{23} &= -2/\sigma_\eta^2 \cdot N_\mathrm{R} h_{1,\mathrm{R}} \sum_p\omega_p  \vect a_{\mathrm{T},1}^T[p] \matt R_s[p] \vect a_{\mathrm{T},1}^\ast[p],\\
	j_{14} &= 2/\sigma_\eta^2 \cdot N_\mathrm{R} \sum_p \Real(h_{1}   \dot{\vect a}_{\mathrm{T},1}^T[p] \matt R_s[p] \vect a_{\mathrm{T},1}^\ast[p]),\\
	j_{24} &= 2/\sigma_\eta^2 \cdot N_\mathrm{R} \sum_p \Imag(h_{1}   \dot{\vect a}_{\mathrm{T},1}^T[p] \matt R_s[p] \vect a_{\mathrm{T},1}^\ast[p]).
\end{align}
\hbadness=7800
\hyphenation{ICCWORKSHOPS}
\bibliography{bistaticsensing}
\end{document}